# Evaluating university research: same performance indicator, different rankings[1]

Giovanni Abramo*
*Laboratory for Studies of Research and Technology Transfer*
*at the Institute for System Analysis and Computer Science (IASI-CNR)*
*National Research Council of Italy*
ADDRESS: Istituto di Analisi dei Sistemi e Informatica, Consiglio Nazionale delle Ricerche, Via dei Taurini 19, 00185 Roma - ITALY
tel. +39 06 7716417, fax +39 06 7716461, giovanni.abramo@uniroma2.it

Ciriaco Andrea D'Angelo
*University of Rome "Tor Vergata" - Italy and*
*Laboratory for Studies of Research and Technology Transfer (IASI-CNR)*
ADDRESS: Dipartimento di Ingegneria dell'Impresa, Università degli Studi di Roma "Tor Vergata", Via del Politecnico 1, 00133 Roma - ITALY
tel. and fax +39 06 72597362, dangelo@dii.uniroma2.it

**Abstract**
Assessing the research performance of multi-disciplinary institutions, where scientists belong to many fields, requires that the evaluators plan how to aggregate the performance measures of the various fields. Two methods of aggregation are possible. These are based on: a) the performance of the individual scientists or b) the performance of the scientific fields present in the institution. The appropriate choice depends on the evaluation context and the objectives for the particular measure. The two methods bring about differences in both the performance scores and rankings. We quantify these differences through observation of the 2008-2012 scientific production of the entire research staff employed in the hard sciences in Italian universities (over 35,000 professors). Evaluators preparing an exercise must comprehend the differences illustrated, in order to correctly select the methodologies that will achieve the evaluation objectives.

---

---

\* *Corresponding author*

# 1. Introduction

The Humboldtian philosophy of "education through research" has forged the higher education systems of many nations. Although questioned in the past, in the face of the massification of higher education, with the growing of more vocationally oriented higher education institutions (Ash, 1999), in the current knowledge-based economy the Humboldtian legacy has regained new attention among policy makers, in terms of the "research–teaching nexus". According to a European report (Commission of the European Communities, 2002), competency-based higher education that is focused on employability in the knowledge society is in need of "education through research". This is because research competencies are useful for professionals in a knowledge society, and because higher education is only able to deliver these competencies if its education is related to research.

The evaluation of university research performance then is becoming ever more common in many nations. The issue of evaluation, as well as the university rankings that are readily available, now attracts the attention of the popular media and a vast and varied public. The *SCImago Institutions Rankings* (SCIMAGO, 2014) and the *CWTS Leiden Rankings* (CWTS, 2014) are the most read and accredited examples of the world rankings carried out by bibliometricians. Other yearly world university rankings attract much greater media and public attention (THE, 2014; QS, 2014; SJTU, 2014), however most bibliometricians agree in strongly criticizing their methodological weaknesses (Billaut et al., 2010; Dehon et al., 2010; Sauder and Espeland, 2009; Liu and Cheng, 2005; Van Raan, 2005).

A growing number of countries now also conduct their own periodical evaluation exercises of research quality in the national university systems. The objectives are multiple. In many nations the results of the evaluations serve in the allocation of public resources (Hicks, 2012; OECD, 2010; Hicks, 2009). The evaluations in fact can have a significant impact on the individual and collective behavior of the actors in the research system (Vanecek, 2014; Himanen et al., 2009; Smart, 2009). Also, where the rankings of the university organizational levels are made public there is a reduction in information asymmetry between the suppliers and the seekers of "new" knowledge, with gains for the efficiency of markets in knowledge and education. Further, students can make informed choices in selecting the institutions for their studies. Private companies can efficiently select partners for joint research, as well as recruit new personnel on the basis of the performance of the universities that provided the candidate training. Universities themselves want to know the strengths and weaknesses of their own organizational units, for purposes of strategic planning. Given such varying motivations, every stakeholder would clearly adopt a unique evaluation perspective, assigning different weights to the dimensions of a performance evaluation as these are linked to the objectives of their concern.

For the state, the typical rationale in allocating more or less state funds to the differently performing universities is to maximize the rate of return on research spending, in the form of yields in scientific and technological advancement. Also, in the light of the above noted research-teaching nexus, increasing the funds to universities that are better in research should translate into overall improvement in the educational offer, and so in the quality of the future labor force. On the basis of the published rankings, students can make informed choices about where to apply for their education. Universities are stimulated to improve performance and rise in the rankings, to attract the best students. The overall process can be one of a virtuous circle, leading to broader economic and social progress.

The objective of this work is to unveil the ambivalence inherent in constructing university research performance ranking lists commissioned by government for efficient funding selection. There are in fact two possible perspectives in facing the problem of maximizing the rate of return on research spending. One perspective is more research-oriented, the other more teaching-oriented. The two perspectives imply different approaches in constructing ranking lists, which presumably lead to different results.

To better exemplify our thinking, consider two universities A and B, identical in size, in fields of research and degree programs, but different in research performance, which is greater in A. Then suppose that the better performance of A is essentially due to 10% of its research staff, which is exceptionally good, even though the remaining 90% have below-median performance in their research fields. Even though all the professors of B have research performance above median, it ranks below A. Now we put ourselves in the shoes of the policy maker that wants to maximize the returns on financing for research. He or she sees these universities as black boxes, a bit like the investor choosing between different stock portfolios. Under parity of conditions, the investor will choose the portfolio offering maximum return, independent of the performance of the individual stocks. Form the research-oriented perspective, the policy maker is also interested in overall returns, rather than the distribution of performance by the individual scientists. The optimal choice will be to allocate resources to university A. From the teaching-oriented perspective, the optimizing choices are not so obvious: is it better to have 10% of classes taught by the "greats" in the subject, and 90% by mediocre professors, or is it better to have fair to good professors for every field of study? The second option could be safer, for a good overall education. Thus from a teaching-oriented perspective, a more penetrating type of ranking seems appropriate and the optimal choice would be to allocate resources to university B.

The extreme institutional characteristics in the above illustration could seem unrealistic, particularly for strongly competitive higher-education systems. However, they could be quite close to reality for systems where competition is largely lacking, as in a number of European nations (Auranen and Nieminen, 2010). For example, a study by Abramo et al. (2012a) has shown that in the Italian case there is a huge dispersion of research performance within the individual universities, compared to what is seen between them. The methods and indicators of evaluation must take account of such potential realities, and be conceived to achieve the objectives of the given evaluation exercise, whether it be for the use of students, in prioritizing funding, or other.

These considerations lead us to examine two of the different methods for measuring the universities' research performance: one that controls for the impact of outliers on the aggregate performance of the institution, and another that does not provide such control. The former would seem to serve better to inform students in the selection of their universities; the second probably results as better to inform decisions aimed at maximizing overall scientific advancements. To rank universities in research performance for a particular field the bibliometrician can proceed either way. The first, based on the performance of individual professors, interprets the performance of the organizational unit as the average of the individual performances, meaning that the emphasis is on the individual. The other method interprets the field as a black box. It normalizes the output of the all the scientists in a field by the labor input, meaning that emphasis is on the overall product of the scientists in that field, independent of the variability of the individual contributions. The two methods, equally legitimate in an operational sense, give rise to performance scores and rankings that are correlated, but still different. The appropriate choice of method then depends on the aims of the evaluation.

The objective of the current work is to consider the implications by quantifying the differences in the rankings by the two methods. To do this we apply the two methodologies to measure the research performance by Italian universities in the hard sciences, at three levels: by field (i.e. radiology, cardiology, algebra, etc.); by discipline (i.e. medicine, mathematics, etc.); and for the overall university. We measure research performance by the productivity indicator Fractional Scientific Strength (see next section), which requires input data. We are aware though that in most countries input data are not available on a large scale. For this reason, we also employ the *new crown indicator* (Waltman et al., 2011).

The next section illustrates the dataset used for the analysis, as well as the indicators of research performance and the calculations for the two methods of preparing the rankings. In Section 3 we compare the performance ranking lists deriving from the two approaches, at the levels of the fields and disciplines. The paper closes with a summary of the findings and some considerations by the authors.

## 2. Data and Methods

### 2.1 Dataset

The dataset for the analysis is the 2008-2012 scientific production achieved by all Italian university professors in the hard sciences. The Italian Ministry of Education, Universities and Research (MIUR) recognizes a total of 96 universities authorized to grant legally recognized degrees. In Italy there are no "teaching-only" universities, as all professors are required to carry out both research and teaching, in keeping with the Humboldtian philosophy of higher education. At 31/12/2013 the entire national university population consisted of 56,600 professors. Each of them is officially classified in one and only one research field. There are a total of 370 such fields (named scientific disciplinary sectors, or SDS[2]), grouped into 14 disciplines (named university disciplinary areas, or UDAs).

It has been shown (Moed, 2005) that in the so-called hard sciences, the prevalent form of codification for research output is publication in scientific journals. For reasons of robustness, we thus examine only the nine UDAs that deal with the hard sciences,[3] including a total of 192 SDSs. Furthermore, again for robustness, we exclude all professors who have been on staff less than three years in the observed period.

Data on academics are extracted from a database maintained at the central level by the MIUR,[4] indexing the name, academic rank, affiliation, and the SDS of each professor. Publication data are drawn from the Italian Observatory of Public Research (ORP), a database developed and maintained by the authors and derived under license from the Web of Science (WoS). Beginning from the raw data of Italian publications[5] indexed in WoS-ORP, we apply a complex algorithm for disambiguation of the true identity of the authors and their institutional affiliations (for details see D'Angelo et al., 2011). Each publication is attributed to the university professors that authored it, with a harmonic average of precision and recall (F-measure) equal to 96 (error of 4%). We further reduce this error by manual disambiguation.

---

[2] The complete list is accessible at http://attiministeriali.miur.it/UserFiles/115.htm, last accessed 17/02/2015.
[3] Mathematics and computer sciences; Physics; Chemistry; Earth sciences; Biology; Medicine; Agricultural and veterinary sciences; Civil engineering; Industrial and information engineering.
[4] http://cercauniversita.cineca.it/php5/docenti/cerca.php, last accessed 17/02/2015.

The dataset for the analysis includes 36,450 scientists, employed in 86 universities, authoring over 200,000 WoS publications, sorted in the UDAs as shown in Table 1.

*Table 1: Dataset for the analysis: number of fields (SDSs), universities, research staff and WoS publications (2008-2012) in each UDA under investigation*

| UDA | SDS | Universities | Research staff | Publications* |
|---|---|---|---|---|
| Mathematics and computer science | 10 | 69 | 3,387 | 16,920 |
| Physics | 8 | 64 | 2,497 | 23,587 |
| Chemistry | 12 | 61 | 3,174 | 26,703 |
| Earth sciences | 12 | 47 | 1,199 | 6,148 |
| Biology | 19 | 66 | 5,198 | 34,399 |
| Medicine | 50 | 64 | 10,966 | 71,575 |
| Agricultural and veterinary sciences | 30 | 55 | 3,207 | 14,209 |
| Civil engineering | 9 | 53 | 1,583 | 6,908 |
| Industrial and information engineering | 42 | 73 | 5,239 | 40,246 |
| Total | 192 | 86 | 36,450 | 206,433† |

\* The figure refers to publications (2008-2012) authored by at least one professor pertaining to the UDA.
† The total is less than the sum of the column data due to double counts of individual publications that pertain to the SDSs of more than one UDA.

**2.2 Measuring research performance**

Research organizations can be likened to any other productive organization, except that rather than producing widgets they engage in producing new knowledge. The principle indicator of a university's production efficiency is in fact its productivity, meaning the ratio between the values of the outputs produced and the inputs used. The measurement of productivity requires certain assumptions and approximations.

As a proxy of output, bibliometricians use the publications indexed in databases such as WoS or Scopus. As proxy of the publications' value (impact), they use the citations. Since citation behavior varies across scientific fields, citations are then normalized by field. However the intensity of publication also varies across fields (Butler, 2007; Moed et al., 1985; Garfield, 1979). To avoid distortions (Abramo et al., 2008) the researchers must be classified in their respective fields, with their performance then normalized by a field-specific scaling factor. The input factors are labor and capital, however capital is not generally known, for which we assume that it is equal for all. What we measure then is not total factor productivity, rather the labor productivity, with this being founded on a strong approximation. The input value is given by the scientists' salaries, where these are known. The indicator we apply to approximate the measure of labor productivity is called *Fractional Scientific Strength* (FSS). The reader is referred to Abramo and D'Angelo (2014) for a more detailed explication of the theory underlying this indicator. In the next subsections we present the two operational methods for the calculation of labor productivity: at the level of the field, discipline and entire institution.

---
[5] We exclude those document types that cannot be strictly considered as true research products, such as editorial material, meeting abstracts, replies to letters, etc.

**2.2.1 University performance based on individual productivity**

The first method for evaluating institutional performance is based on the measure of productivity of the individual scientists. At the level of in the individual professor *P*, the average yearly productivity over a period of time, accounting for the cost of labor is:

$$FSS_P = \frac{1}{w_P} \cdot \frac{1}{t} \sum_{i=1}^{N} \frac{c_i}{\bar{c}} f_i$$

[1]

Where:
$w_P$ = average yearly salary of the professor;[6]
t = number of years the professor worked over the period of observation;
N = number of publications by the professor over the period of observation;
$c_i$ = citations received for publication *i* (observed at May 15, 2014);
$\bar{c}$ = average of the distribution of citations received for all cited publications[7] indexed in the same year and subject category as publication *i*;
$f_i$ = fractional contribution of the professor to publication *i*.

Fractional contribution equals the inverse of the number of authors, in those fields where the practice is to place the authors in simple alphabetical order, but assumes different weights in other cases. For the life sciences, widespread practice in Italy and abroad is for the authors to indicate the various contributions to the published research by the order of the names in the byline. For these areas, we give different weights to each co-author according to their order in the byline and the character of the co-authorship (intra-mural or extra-mural). If first and last authors belong to the same university, 40% of citations are attributed to each of them; the remaining 20% are divided among all other authors. If the first two and last two authors belong to different universities, 30% of citations are attributed to first and last authors; 15% of citations are attributed to second and last author but one; the remaining 10% are divided among all others[8]. Failure to account for the number and position of authors in the byline would result in notable ranking distortions at both the individual (Abramo et al. 2013a) and aggregate (Abramo et al. 2013b) levels.

The first of the two methods for measurement of university productivity in a field, discipline or "overall" involves standardization of individual productivity by the SDS average. In formula, the productivity $FSS_{U_P}$ over a certain period for university *U*, in a field, discipline and overall is:

---

[6] This information is unavailable for reasons of privacy. We resort to a proxy, i.e. the nationally averaged salary of the professors in each academic rank (data source DALIA – MIUR, https://dalia.cineca.it/php4/inizio_access_cnvsu.php, last accessed 17/02/2015). Failure to account for the cost of labor would result in ranking distortions, as shown by Abramo et al. (2010).
[7] For details about the choice of this scaling factor, see Abramo et al. (2012b).
[8] The weighting values were assigned following advice from senior Italian professors in the life sciences. The values could be changed to suit different practices in other national contexts.

$$FSS_{U_P} = \frac{1}{RS}\sum_{j=1}^{RS}\frac{FSS_{P_j}}{\overline{FSS_P}}$$

[2]

Where:
$RS$ = research staff of the field/discipline/institution, in the observed period;
$FSS_{P_j}$ = productivity of professor $j$;
$\overline{FSS_P}$ = national average productivity of all productive professors in the same SDS as professor $j$.

An alternative approach would be to express the productivity at the aggregate levels by the simple average of the percentile ranks of the researchers. However this method is obviously subject to limitations, the first being the compression of the performance differences between one position and the next. Thompson (1993) warns that percentile ranks should not be added or averaged, because percentile is a numeral that does not represent equal-interval measurement. Further, percentile rank is also sensitive to the size of the fields and to the performance distribution.

**2.2.2 University performance based on SDS productivity**

The productivity of an institution at the field (SDS) level can alternatively be calculated considering the SDS as a black box. In other words we measure the entire output of the professors in the SDS, without distinguishing the individuals' production. the yearly average productivity $FSS_S$ of an institution in SDS $S$, over a certain period is:

$$FSS_S = \frac{1}{w_S}\sum_{i=1}^{N}\frac{c_i}{\bar{c}}f_i$$

[3]

Where:
$w_S$ = total salary of the professors in the SDS at the given university, over the observed period;
$N$ = number of publications by professors in the same SDS over the observed period;
$c_i$ = citations received for publication $i$;
$\bar{c}$ = average citations received for all cited publications indexed in the same year and subject category of publication $i$;
$f_i$ = fractional contribution of the professors in the SDS to publication $i$, calculated as described above.

The performance evaluation of an institution at higher aggregate levels (UDA, overall) requires standardization of the different SDSs' productivity by the national averages, and weighting by SDS size. In formula, the productivity $FSS_{U_S}$ over a certain period for UDA $U$ of a university, is:

$$FSS_{U_S} = \sum_{k=1}^{N_U}\frac{FSS_{S_k}}{\overline{FSS_{S_k}}}\frac{w_{S_k}}{w_U}$$

[4]

With:
$w_{S_k}$ = total salary of the professors in SDS $k$ at the given university, over the observed

period;
$w_U$ = total salary of professors in UDA $U$ at the university, over the observed period;
$N_U$ = number of SDSs in UDA $U$ at the given university;
$\overline{FSS_{S_k}}$ = weighted[9] average $FSS_S$ of all universities with productivity above 0, in the SDS $k$.

## 3. Results

The two methods of calculating research performance give rise to different scores, and presumably different rankings. In this section we will compare the rankings for the case of Italian universities in the hard sciences, beginning from the SDS level and proceeding to greater aggregation.

### 3.1 Comparing university productivity at the SDS level

We calculate the productivity at the SDS level using the two methods: aggregating and averaging the measures for the individual researchers ($FSS_P$), and calculating the ratio of the SDS output to the total salary of its professors ($FSS_S$). Table 2 provides the example of the results from the measures for the Italian universities active in Nuclear and subnuclear physics (SDS FIS/04). Columns three and six show the values of the two measures for the 25 universities active in this SDS. The rank correlation under the two methods is remarkably high (Spearman ρ = 0.960), but important differences are also evident. In particular, three-quarters of the universities change position, with an average shift of 1.52 positions and the maximum shift at UNIV_16, climbing from 16th for $FSS_P$ to 11th place for $FSS_S$.

Figure 1 gives a visual indication of the shifts in rank for the individual universities, evaluated under the two methods. We see significant volatility, including at the top of the ranking, where UNIV_1 and 3 switch first place, as well as a concentration of shifts in the area from 15th to 21st position.

Table 3 broadens the statistics to the other SDSs in Physics, giving an idea of the variability across fields in a single discipline. We see that FIS/08 (Didactics and history of phyiscs) is the only field where the two rankings are perfectly overlapping. In the others, the variations of rank concern a minimum of 54.5% (FIS/06-Physics for earth and atmospheric sciences), varying up to 85.3% of the universities in the field (FIS/02-Theoretical physics, mathematical models and methods). The FIS/02 SDS is clearly the field featuring the most substantial average shifts (7.7 percentile points), plus it has the extreme case of one of the 34 universities shifting a full 12 positions (36.4 percentile points), going from 29th place under $FSS_P$ to 17th by $FSS_S$.

---

[9] The weighting accounts for the relative size (in terms of cost of labor) of the SDSs of each university.

*Table 2: Productivity scores and rankings by $FSS_P$ and $FSS_S$ for Italian universities in FIS/04 (Nuclear and subnuclear physics)*

|  |  | $FSS_P$ |  |  | $FSS_S$ |  |  | Rank shift | Percentile shift |
|---|---|---|---|---|---|---|---|---|---|
| ID* | Research staff | value | rank | percentile | value | rank | percentile |  |  |
| UNIV_1 | 6 | 0.664 | 1 | 100.0 | 0.104 | 3 | 91.7 | ↓ 2 | - 8.3 |
| UNIV_2 | 2 | 0.550 | 2 | 95.8 | 0.098 | 4 | 87.5 | ↓ 2 | - 8.3 |
| UNIV_3 | 2 | 0.540 | 3 | 91.7 | 0.120 | 1 | 100 | ↑ 2 | + 8.3 |
| UNIV_4 | 10 | 0.496 | 4 | 87.5 | 0.086 | 5 | 83.3 | ↓ 1 | - 4.2 |
| UNIV_5 | 12 | 0.476 | 5 | 83.3 | 0.105 | 2 | 95.8 | ↑ 3 | + 12.5 |
| UNIV_6 | 6 | 0.414 | 6 | 79.2 | 0.072 | 7 | 75.0 | ↓ 1 | - 4.2 |
| UNIV_7 | 5 | 0.353 | 7 | 75.0 | 0.074 | 6 | 79.2 | ↑ 1 | + 4.2 |
| UNIV_8 | 7 | 0.340 | 8 | 70.8 | 0.068 | 8 | 70.8 | = | = |
| UNIV_9 | 6 | 0.332 | 9 | 66.7 | 0.062 | 10 | 62.5 | ↓ 1 | - 4.2 |
| UNIV_10 | 3 | 0.235 | 10 | 62.5 | 0.063 | 9 | 66.7 | ↑ 1 | + 4.2 |
| UNIV_11 | 7 | 0.232 | 11 | 58.3 | 0.052 | 12 | 54.2 | ↓ 1 | - 4.2 |
| UNIV_12 | 8 | 0.232 | 12 | 54.2 | 0.044 | 14 | 45.8 | ↓ 2 | - 8.3 |
| UNIV_13 | 3 | 0.232 | 13 | 50.0 | 0.044 | 15 | 41.7 | ↓ 2 | - 8.3 |
| UNIV_14 | 10 | 0.205 | 14 | 45.8 | 0.049 | 13 | 50.0 | ↑ 1 | + 4.2 |
| UNIV_15 | 3 | 0.194 | 15 | 41.7 | 0.034 | 18 | 29.2 | ↓ 3 | - 12.5 |
| UNIV_16 | 6 | 0.192 | 16 | 37.5 | 0.054 | 11 | 58.3 | ↑ 5 | + 20.8 |
| UNIV_17 | 8 | 0.182 | 17 | 33.3 | 0.040 | 16 | 37.5 | ↑ 1 | + 4.2 |
| UNIV_18 | 6 | 0.173 | 18 | 29.2 | 0.026 | 22 | 12.5 | ↓ 4 | - 16.7 |
| UNIV_19 | 2 | 0.163 | 19 | 25.0 | 0.033 | 19 | 25.0 | = | = |
| UNIV_20 | 4 | 0.156 | 20 | 20.8 | 0.027 | 20 | 20.8 | = | = |
| UNIV_21 | 6 | 0.151 | 21 | 16.7 | 0.039 | 17 | 33.3 | ↑ 4 | + 16.7 |
| UNIV_22 | 5 | 0.144 | 22 | 12.5 | 0.027 | 21 | 16.7 | ↑ 1 | + 4.2 |
| UNIV_23 | 3 | 0.055 | 23 | 8.3 | 0.019 | 23 | 8.3 | = | = |
| UNIV_24 | 5 | 0.051 | 24 | 4.2 | 0.011 | 24 | 4.2 | = | = |
| UNIV_25 | 2 | 0.035 | 25 | 0 | 0.007 | 25 | 0 | = | = |

*\* The population consists of universities (25 in all) having at least 2 professors in the SDS*

*Figure 1: University productivity rankings (percentile) by $FSS_P$ and $FSS_S$ in FIS/04-Nuclear and subnuclear physics*

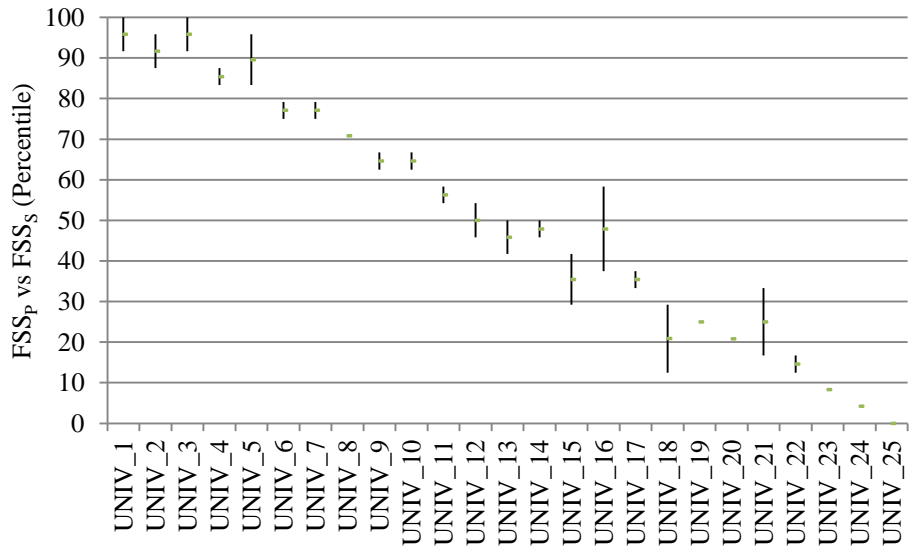

*Table 3: Comparison of productivity rankings by $FSS_P$ and $FSS_S$ in the SDSs of Physics (percentile shift in brackets)*

| SSD | Universities* | Spearman ρ | % shifting | Average shift | Median shift | Max shift |
|---|---|---|---|---|---|---|
| FIS/01 | 51 | 0.978 | 84.3 | 2.20 (4.4) | 2 | 10 (20.0) |
| FIS/02 | 34 | 0.930 | 85.3 | 2.53 (7.7) | 1.5 | 12 (36.4) |
| FIS/03 | 38 | 0.982 | 81.6 | 1.53 (4.1) | 1 | 5 (13.5) |
| FIS/04 | 25 | 0.960 | 76.0 | 1.52 (6.3) | 1 | 5 (20.8) |
| FIS/05 | 23 | 0.979 | 69.6 | 1.04 (4.7) | 1 | 3 (13.6) |
| FIS/06 | 11 | 0.973 | 54.5 | 0.55 (5.5) | 1 | 1 (10.0) |
| FIS/07 | 41 | 0.979 | 68.3 | 1.61 (4.0) | 1 | 8 (20.0) |
| FIS/08 | 10 | 1 | 0.0 | 0.00 (0.0) | 0 | 0 (0) |

*\* The population consists of universities having at least 2 professors in the SDS*

Continuing with the Italian case, this same type of analysis is extended to all the 192 national SDSs investigated, as summarized in Table 4. The coefficients of correlation between the two rankings drop below 0.9 in only two fields: BIO/08-Anthropology (0.888) and MED/37-Neuroradiology (0.867). The rankings superimpose perfectly in 17 of the SDSs. Other than FIS/06 of Physics, this occurs in one SDS of Earth sciences, one of Medicine, three of Agricultural and veterinary sciences and 11 of Industrial and information engineering. However in every UDA there are SDSs with remarkable variations in positioning between the two rankings, both in terms of percentage of universities shifting rank (maximums are never less than 76%), in average shift (maximums are never less than 5 percentile points), and in max shift (maximums are never less than 21 percentiles).

*Table 4: Comparison of productivity rankings by $FSS_P$ and $FSS_S$ (min-max of percentile variations) for the SDSs of each UDA*

| UDA* | No. of SDSs | Range of variation (min-max) of universities experiencing shift (%) | Range of variation (min-max) of average shift (percentiles) | Range of variation (min-max) in max shift (percentiles) | Spearman ρ (min-max) |
|---|---|---|---|---|---|
| 1 | 10 | (27.3-86.5) | (1.3-7.6) | (4.8-29.3) | (0.917-0.994) |
| 2 | 8 | (0.0-85.3) | (0-7.7) | (0-36.4) | (0.930-1) |
| 3 | 12 | (25.0-82.8) | (3.6-6.6) | (12.2-30.8) | (0.938-0.984) |
| 4 | 12 | (0.0-80.0) | (0-7.8) | (0-41.2) | (0.903-1) |
| 5 | 19 | (36.0-86.0) | (2.3-8.3) | (11.5-40.0) | (0.888-0.990) |
| 6 | 50 | (0.0-83.3) | (0-11.1) | (0-37.5) | (0.867-1) |
| 7 | 30 | (0.0-76.9) | (0-10.3) | (0-30.8) | (0.918-1) |
| 8 | 9 | (30.4-76.7) | (2.4-5.2) | (13.6-21.1) | (0.971-0.986) |
| 9 | 42 | (0.0-84.8) | (0-7.4) | (0-28.6) | (0.943-1) |

*\* 1 = Mathematics and computer sciences; 2 = Physics; 3 = Chemistry; 4 = Earth sciences; 5 = Biology; 6 = Medicine; 7 = Agricultural and veterinary sciences; 8 = Civil engineering; 9 = Industrial and information engineering.*

### 3.2 Comparing university productivity at the UDA level

The above analyses show that, at the SDS level, there are very strong correlations between the rankings under the two different methods. However equally clearly, there are also very meaningful differences in rank for some universities. In this section we repeat the analysis for the higher levels of aggregation: the UDA and the overall university. Table 5 presents the rankings in Physics calculated with formulae [2] and [4] for the 43 universities with research staff of at least 10 professors.

The correlation between the rankings under the two approaches is very strong (Spearman ρ = 0.984). There are 12 universities where the rank remains constant, including the first three and last five. The universities with middle ranks are the ones that experience the greatest shifts, with the maximum being 7 positions (17 percentile points).

The frequency distribution of the rank shifts (Table 5, second-last column) is diagrammed in Figure 2. The mode of the left-skewed shift distribution is nil.

*Table 5: Productivity scores and rankings by $FSS_{U_P}$ and $FSS_{U_S}$ for Italian universities in Physics*

| ID* | Res. staff | $FSS_{U_P}$ value | rank* | percentile | $FSS_{U_S}$ value | rank* | percentile | Rank shift | Percentile shift |
|---|---|---|---|---|---|---|---|---|---|
| UNIV_26 | 17 | 2.165 | 1 | 100.0 | 2.625 | 1 | 100.0 | = | 0 |
| UNIV_27 | 54 | 2.130 | 2 | 97.6 | 2.389 | 2 | 97.6 | = | 0 |
| UNIV_28 | 15 | 1.657 | 3 | 95.2 | 1.740 | 3 | 95.2 | = | 0 |
| UNIV_29 | 15 | 1.608 | 4 | 92.9 | 1.679 | 5 | 90.5 | ↓ 1 | - 2.4 |
| UNIV_30 | 31 | 1.477 | 5 | 90.5 | 1.688 | 4 | 92.9 | ↑ 1 | + 2.4 |
| UNIV_31 | 51 | 1.343 | 6 | 88.1 | 1.363 | 7 | 85.7 | ↓ 1 | - 2.4 |
| UNIV_1 | 39 | 1.314 | 7 | 85.7 | 1.463 | 6 | 88.1 | ↑ 1 | + 2.4 |
| UNIV_32 | 16 | 1.232 | 8 | 83.3 | 1.204 | 10 | 78.6 | ↓ 2 | - 4.7 |
| UNIV_33 | 20 | 1.145 | 9 | 81.0 | 1.348 | 8 | 83.3 | ↑ 1 | + 2.3 |
| UNIV_9 | 73 | 1.105 | 10 | 78.6 | 1.190 | 11 | 76.2 | ↓ 1 | - 2.4 |
| UNIV_2 | 93 | 1.087 | 11 | 76.2 | 1.175 | 13 | 71.4 | ↓ 2 | - 4.8 |
| UNIV_5 | 92 | 1.078 | 12 | 73.8 | 1.185 | 12 | 73.8 | = | 0 |
| UNIV_4 | 66 | 1.059 | 13 | 71.4 | 1.209 | 9 | 81.0 | ↑ 4 | + 9.6 |
| UNIV_16 | 162 | 0.997 | 14 | 69.0 | 1.150 | 14 | 69.0 | = | 0 |
| UNIV_6 | 86 | 0.993 | 15 | 66.7 | 1.048 | 18 | 59.5 | ↓ 3 | - 7.2 |
| UNIV_12 | 126 | 0.981 | 16 | 64.3 | 1.128 | 15 | 66.7 | ↑ 1 | + 2.4 |
| UNIV_13 | 27 | 0.970 | 17 | 61.9 | 1.069 | 16 | 64.3 | ↑ 1 | + 2.4 |
| UNIV_10 | 39 | 0.970 | 18 | 59.5 | 1.048 | 17 | 61.9 | ↑ 1 | + 2.4 |
| UNIV_19 | 59 | 0.951 | 19 | 57.1 | 1.010 | 21 | 52.4 | ↓ 2 | - 4.7 |
| UNIV_34 | 12 | 0.923 | 20 | 54.8 | 0.893 | 27 | 38.1 | ↓ 7 | - 16.7 |
| UNIV_35 | 19 | 0.918 | 21 | 52.4 | 1.037 | 20 | 54.8 | ↑ 1 | + 2.4 |
| UNIV_24 | 120 | 0.898 | 22 | 50.0 | 1.040 | 19 | 57.1 | ↑ 3 | + 7.1 |
| UNIV_22 | 68 | 0.877 | 23 | 47.6 | 0.971 | 22 | 50.0 | ↑ 1 | + 2.4 |
| UNIV_18 | 72 | 0.854 | 24 | 45.2 | 0.881 | 28 | 35.7 | ↓ 4 | - 9.5 |
| UNIV_3 | 47 | 0.853 | 25 | 42.9 | 0.943 | 24 | 45.2 | ↑ 1 | + 2.3 |
| UNIV_7 | 53 | 0.830 | 26 | 40.5 | 0.963 | 23 | 47.6 | ↑ 3 | + 7.1 |
| UNIV_36 | 15 | 0.813 | 27 | 38.1 | 0.732 | 33 | 23.8 | ↓ 6 | - 14.3 |
| UNIV_11 | 165 | 0.779 | 28 | 35.7 | 0.901 | 26 | 40.5 | ↑ 2 | + 4.8 |
| UNIV_37 | 66 | 0.768 | 29 | 33.3 | 0.910 | 25 | 42.9 | ↑ 4 | + 9.6 |
| UNIV_21 | 53 | 0.766 | 30 | 31.0 | 0.851 | 29 | 33.3 | ↑ 1 | + 2.3 |
| UNIV_8 | 98 | 0.751 | 31 | 28.6 | 0.783 | 31 | 28.6 | = | 0 |
| UNIV_38 | 14 | 0.738 | 32 | 26.2 | 0.812 | 30 | 31.0 | ↑ 2 | + 4.8 |
| UNIV_39 | 44 | 0.700 | 33 | 23.8 | 0.715 | 34 | 21.4 | ↓ 1 | - 2.4 |
| UNIV_14 | 89 | 0.685 | 34 | 21.4 | 0.762 | 32 | 26.2 | ↑ 2 | + 4.8 |
| UNIV_40 | 16 | 0.628 | 35 | 19.0 | 0.664 | 38 | 11.9 | ↓ 3 | - 7.1 |
| UNIV_41 | 39 | 0.601 | 36 | 16.7 | 0.714 | 35 | 19.0 | ↑ 1 | + 2.3 |
| UNIV_17 | 97 | 0.583 | 37 | 14.3 | 0.664 | 37 | 14.3 | = | 0 |
| UNIV_20 | 41 | 0.573 | 38 | 11.9 | 0.708 | 36 | 16.7 | ↑ 2 | + 4.8 |
| UNIV_25 | 50 | 0.510 | 39 | 9.5 | 0.633 | 39 | 9.5 | = | 0 |
| UNIV_23 | 43 | 0.504 | 40 | 7.1 | 0.566 | 40 | 7.1 | = | 0 |
| UNIV_42 | 57 | 0.495 | 41 | 4.8 | 0.557 | 41 | 4.8 | = | 0 |
| UNIV_15 | 48 | 0.489 | 42 | 2.4 | 0.543 | 42 | 2.4 | = | 0 |
| UNIV_43 | 15 | 0.443 | 43 | 0.0 | 0.458 | 43 | 0.0 | = | 0 |

*\* The population consists of universities (43 in all) having at least 10 professors in the UDA*

*Figure 2: Frequency distribution of university rank shifts in Physics*

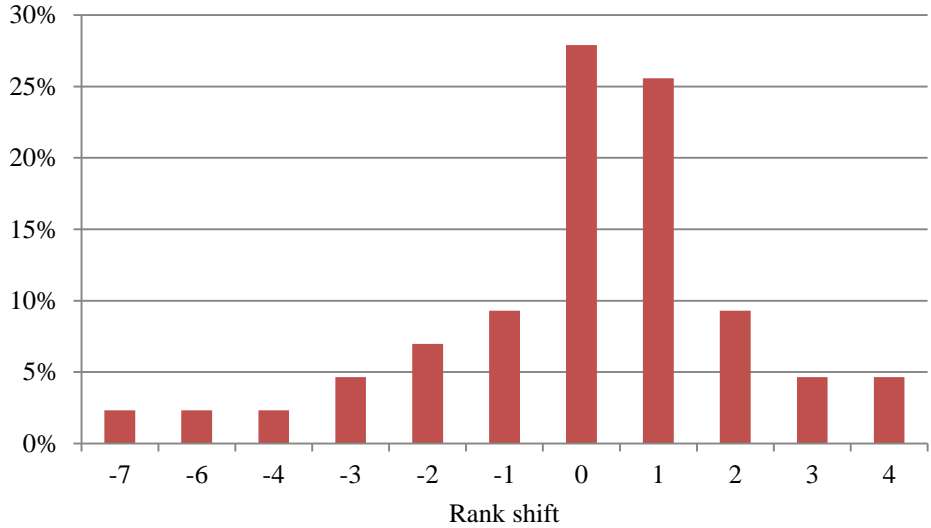

We repeat the above analysis for all nine disciplines (Italian UDAs) observed, at the level of the entire university (Table 6). Overall, the variations between ranks by $FSS_{U_P}$ and by $FSS_{U_S}$ affect 67.2% of the universities evaluated, with a maximum of 90.4% in Biology and a minimum of 59.4% in Earth sciences. The average rank shift for the universities under the two evaluation methods is least in Earth sciences, at 1.1 places (3.4 percentiles), and greatest in Mathematics, at 2.7 places (5.6 percentiles). However in the case of percentile shifts, the Agricultural and veterinary science UDA shows the highest average movement (6.2). The UDAs of Industrial and information engineering and Mathematics show the individual cases of the greatest rank shifts, of 12 positions (24.5 and 26.1 percentiles respectively). The median of the rank differences is never zero (between 1 and 2, according to the UDA). The Spearman coefficient is always very high: although it is least in Agricultural and veterinary sciences, even here it is still 0.959.

*Table 6: Comparison of productivity rankings by $FSS_{U_P}$ and $FSS_{U_S}$, in the nine UDAs observed (percentile shift in brackets)*

| UDA* | No. of universities§ | % shifting rank | Average shift | Median shift | Max shift | Correlation |
|---|---|---|---|---|---|---|
| 1 | 50 | 82.0 | 2.7 (5.6) | 2 (4.1) | 12 (24.5) | 0.964 |
| 2 | 43 | 72.1 | 1.5 (3.7) | 1 (2.4) | 7 (16.7) | 0.984 |
| 3 | 44 | 75.0 | 2.5 (5.7) | 2 (4.7) | 9 (20.9) | 0.961 |
| 4 | 32 | 59.4 | 1.1 (3.4) | 1 (3.2) | 4 (12.9) | 0.985 |
| 5 | 52 | 90.4 | 2.1 (4.1) | 2 (3.9) | 9 (17.6) | 0.982 |
| 6 | 43 | 76.7 | 1.8 (4.2) | 1 (2.4) | 8 (19.0) | 0.980 |
| 7 | 29 | 79.3 | 1.7 (6.2) | 1 (3.6) | 7 (25.0) | 0.959 |
| 8 | 36 | 77.8 | 1.3 (3.7) | 1 (2.9) | 4 (11.4) | 0.987 |
| 9 | 47 | 70.2 | 2.1 (4.5) | 2 (4.3) | 12 (26.1) | 0.975 |
| Total | 64† | 67.2 | 2.1 (3.3) | 1 (1.6) | 21 (33.3) | 0.978 |

\* 1 = Mathematics and computer sciences; 2 = Physics; 3 = Chemistry; 4 = Earth sciences; 5 = Biology; 6 = Medicine; 7 = Agricultural and veterinary sciences; 8 = Civil engineering; 9 = Industrial and information engineering.
§ *The population consists of universities having at least 10 professors in the UDA*
† *The population consists of universities having overall at least 30 professors*

Applying formulae [2] and [4] to all the SDSs active in the universities, independent of the UDA, we obtain general scores and ranking lists that present some surprises. The last row of Table 6 indicates that, although there is still very high correlation between the rankings (Spearman ρ = 0.978), we also observe extreme outliers: one university leaps ahead 21 positions, shifting from 41$^{st}$ place under $FSS_{U_P}$ to 20$^{th}$ by $FSS_{U_S}$. Another two universities lose a full nine positions, while another gains eight.

It also seems useful to analyze the variations in rankings by quartile, rather than by absolute ranking: the quartile classification is typically used in national evaluation exercises (Table 7).

*Table 7: Comparison of quartile productivity rankings by $FSS_{U_P}$ and $FSS_{U_S}$, in the nine UDAs observed*

| UDA* | No. of universities§ | % shifting quartile | Average shift | Max shift | Spearman ρ correlation | % from top to non top |
|---|---|---|---|---|---|---|
| 1 | 50 | 16.0 | 0.2 | 1 | 0.938 | 6.0 |
| 2 | 43 | 14.0 | 0.1 | 1 | 0.945 | 2.3 |
| 3 | 44 | 18.2 | 0.2 | 1 | 0.927 | 2.3 |
| 4 | 32 | 12.5 | 0.1 | 1 | 0.950 | 0.0 |
| 5 | 52 | 15.4 | 0.2 | 1 | 0.938 | 1.9 |
| 6 | 43 | 14.0 | 0.1 | 1 | 0.945 | 4.7 |
| 7 | 29 | 6.9 | 0.1 | 1 | 0.973 | 3.4 |
| 8 | 36 | 11.1 | 0.1 | 1 | 0.956 | 2.8 |
| 9 | 47 | 17.0 | 0.2 | 1 | 0.933 | 2.1 |
| Total | 64† | 12.5 | 0.1 | 1 | 0.950 | 1.6 |

*\* 1 = Mathematics and computer sciences; 2 = Physics; 3 = Chemistry; 4 = Earth sciences; 5 = Biology; 6 = Medicine; 7 = Agricultural and veterinary sciences; 8 = Civil engineering; 9 = Industrial and information engineering.*
*§ The population consists of universities having at least 10 professors in the UDA*
*† The population consists of universities having overall at least 30 professors*

The percentage of universities that register a variation of quartile "productivity rank" under the two methods oscillates between 6.9% in Agricultural and veterinary sciences and 18.2% in Chemistry. There are no shifts greater than one quartile. However looking at the crucial top quartile in particular, 6% of the 50 universities evaluated in Mathematics, which were at the top for $FSS_{U_P}$, are no longer "top" when ranked by $FSS_{U_S}$. Analogous situations occur in all the UDAs with the exception of Earth sciences. With each of these variations from the first to second quartile there is obviously an analogous shift in the opposite direction, with universities moving from second to first. Thus, in spite of the high correlation between the results (Spearman ρ never below 0.92), the two methods of evaluation produce unavoidable variations in ranking.

To close, we offer a comparison of the evaluation rankings for the disciplines in just one university, in this case UNIV 12 (Table 8). This is a large, generalist university, with a research staff of 1,540 professors in the nine UDAs examined. We observe that in changing from evaluation by $FSS_{U_P}$ to $FSS_{U_S}$, the position of this university in the overall ranking remains stable (Table 8, last row). However at the level of the UDA, the ranking is unchanged only in Medicine, while it improves in Agricultural and veterinary sciences and Physics and worsens in all the other UDAs. The greatest shift in position occurs in Chemistry.

*Table 8: Productivity scores and national rankings by UDAs in a large generalist university (UNIV 12)*

| UDA* | Research staff | $FSS_{U_P}$ value | rank† | percentile | $FSS_{U_S}$ value | rank† | percentile | Rank shift | Percentile shift |
|---|---|---|---|---|---|---|---|---|---|
| 1 | 135 | 0.683 | 29 of 50 | 42.9 | 0.812 | 33 of 50 | 34.7 | ↓ 4 | - 8.2 |
| 2 | 126 | 0.981 | 16 of 43 | 64.3 | 1.128 | 15 of 43 | 66.7 | ↑ 1 | + 2.4 |
| 3 | 133 | 1.020 | 13 of 44 | 72.1 | 1.000 | 18 of 44 | 60.5 | ↓ 5 | - 11.6 |
| 4 | 52 | 1.522 | 2 of 32 | 96.8 | 1.608 | 3 of 32 | 93.5 | ↓ 1 | - 3.2 |
| 5 | 205 | 1.243 | 5 of 52 | 92.2 | 1.250 | 6 of 52 | 90.2 | ↓ 1 | - 2.0 |
| 6 | 406 | 1.410 | 4 of 43 | 92.9 | 1.523 | 4 of 43 | 92.9 | = | 0 |
| 7 | 189 | 1.272 | 3 of 29 | 92.9 | 1.739 | 2 of 29 | 96.4 | ↑ 1 | + 3.6 |
| 8 | 57 | 1.050 | 5 of 36 | 88.6 | 1.298 | 6 of 36 | 85.7 | ↓ 1 | - 2.9 |
| 9 | 237 | 1.189 | 5 of 47 | 91.3 | 1.360 | 6 of 47 | 89.1 | ↓ 1 | - 2.2 |
| Total | | 1.195 | 8 of 64 | 88.9 | 1.339 | 8 of 64 | 88.9 | = | 0 |

\* 1 = Mathematics and computer sciences; 2 = Physics; 3 = Chemistry; 4 = Earth sciences; 5 = Biology; 6 = Medicine; 7 = Agricultural and veterinary sciences; 8 = Civil engineering; 9 = Industrial and information engineering.

† The population consists of universities having at least 10 professors in the UDA

### 3.3 Comparing university performance by the *new crown indicator*

In this section we repeat the same analyses by the *new crown indicator* (Waltman et al., 2011), or mean normalized citation score (MNCS), in place of productivity. We call $MNCS_P$ the *new crown indicator* of each professor and $MNCS_S$ the *new crown indicator* of each SDS. We then construct two university ranking lists at SDS level, one by averaging the $MNCS_P$ of all professors of the SDS, and the other by $MNCS_S$. Table 9 shows few descriptive statistics of the shifts between the two rankings for SDS of each UDA. As compared to Table 4, shifts are more noticeable. In each UDA there is at least one SDS showing a correlation coefficient below 0.7. The absolute minimum correlation ($\rho = 0.154$) occurs in an SDS of UDA 6 (Medicine). In at least one SDS of Mathematics, Chemistry, Biology, and Agricultural and veterinary sciences there are universities shifting 90 percentile points. In AGR/07 (Agrarian genetics), the university of Udine jumps from bottom in rank by average $MNCS_P$ to top by $MNCS_S$. In three UDAs (4-Earth sciences, 6-Medicine, and 9-Industrial and information engineering) there is at least an SDS with no shifts at all, while other SDSs in the same UDAs experience shifts of up to 70 percentile points.

*Table 9: Comparison of university rankings by MNCS for the SDSs of each UDA*

| UDA* | No. of SDSs | Range of variation (min-max) of universities experiencing shift (%) | Range of variation (min-max) of average shift (percentiles) | Range of variation (min-max) in max shift (percentiles) | Spearman ρ (min-max) |
|---|---|---|---|---|---|
| 1 | 10 | (66.7-97.6) | (8.0-20.2) | (27.3-90.5) | (0.212-0.919) |
| 2 | 8 | (30.0-96.1) | (8.9-18.2) | (31.8-67.6) | (0.375-0.880) |
| 3 | 12 | (55.6-96.6) | (9.2-21.4) | (42.9-92.9) | (0.313-0.878) |
| 4 | 12 | (0.0-100.0) | (0-19.0) | (0-73.7) | (0.629-1) |
| 5 | 19 | (56.3-96.7) | (8.3-14.5) | (30.3-96.8) | (0.465-0.909) |
| 6 | 50 | (0.0-100.0) | (0-29.7) | (0-80.6) | (0.154-1) |
| 7 | 30 | (50.0-100.0) | (9.1-33.3) | (26.1-100) | (0.200-0.916) |
| 8 | 9 | (60.9-96.8) | (7.5-22.1) | (29.6-75.0) | (0.372-0.900) |
| 9 | 42 | (0.0-96.0) | (0-33.3) | (0-85.7) | (0.262-1) |

\* 1 = Mathematics and computer sciences; 2 = Physics; 3 = Chemistry; 4 = Earth sciences; 5 = Biology; 6 = Medicine; 7 = Agricultural and veterinary sciences; 8 = Civil engineering; 9 = Industrial and information engineering.

We then constructed university rankings by MNCS at the UDA level. Table 10 shows few descriptive statistics of shifts in rank in each UDA. Overall, the variations between ranks affect 93.8% of the universities evaluated, with a maximum of 100% in Agricultural and veterinary sciences and a minimum of 87.5% in Earth sciences. The average rank shift for the universities under the two evaluation methods is never below 11 percentile points, while the maximum shift (19 percentile points) occurs in Civil engineering. In Industrial and information engineering the 47 universities observed experience on average a shift of 8 positions, while the maximum shift is of 37 positions. Variation in ranking by MNCS is more volatile than by FSS. The Spearman ρ coefficients, although never drop below 0.6 are never above 0.87, while the minimum by FSS was 0.959 (see Table 6). The higher variability between the two ranking lists occurs at overall university level (bottom row of Table 10): Spearman ρ equals 0.580, the average shift is 11.1 positions out of 64 (17.6 percentile points), and the maximum shift is 53 positions.

*Table 10: Comparison of university rankings by MNCS in the nine UDAs observed (percentile shift in brackets)*

| UDA* | No. of universities§ | % shifting rank | Average shift | Median shift | Max shift | Correlation |
|---|---|---|---|---|---|---|
| 1 | 50 | 92.0 | 7.4 (15.2) | 4 (8.1) | 42 (85.7) | 0.684 |
| 2 | 43 | 93.0 | 4.8 (11.5) | 4 (9.5) | 19 (45.2) | 0.870 |
| 3 | 44 | 90.9 | 6.4 (14.9) | 3 (7.0) | 27 (62.8) | 0.752 |
| 4 | 32 | 87.5 | 5.3 (17.1) | 5 (16.1) | 19 (61.3) | 0.703 |
| 5 | 52 | 88.5 | 6.6 (12.9) | 4 (7.8) | 32 (62.7) | 0.808 |
| 6 | 43 | 93.0 | 5.3 (12.7) | 4 (9.5) | 17 (40.5) | 0.836 |
| 7 | 29 | 100.0 | 3.9 (13.8) | 3 (10.7) | 16 (57.1) | 0.814 |
| 8 | 36 | 94.4 | 6.7 (19.0) | 6 (17.1) | 19 (54.3) | 0.662 |
| 9 | 47 | 91.5 | 8.1 (17.6) | 6 (13.0) | 37 (80.4) | 0.657 |
| Total | 64† | 93.8 | 11.1 (17.6) | 7 (11.1) | 53 (84.1) | 0.580 |

*\* 1 = Mathematics and computer sciences; 2 = Physics; 3 = Chemistry; 4 = Earth sciences; 5 = Biology; 6 = Medicine; 7 = Agricultural and veterinary sciences; 8 = Civil engineering; 9 = Industrial and information engineering.*
*§ The population consists of universities having at least 10 professors in the UDA*
*† The population consists of universities having overall at least 30 professors*

## 5. Conclusions

Measuring university research performance can be accomplished with two different methodologies. These lead to scores and ranking lists that are different. One of the methods is based on the average performance of the individual researchers in a field, while the other is based on the overall performance from the field, essentially considering it a "black box". By aggregating the results from either approach, the evaluations can arrive at the average performance of the professors or of the fields, within the larger organizational units and the entire institution.

The two approaches will clearly give results that are correlated, however there could be relevant differences for the single departments, disciplines and institutions. This is exactly what we have observed in the current study, based on the observation of the 2008-2012 scientific production of the entire academic staff employed in the hard sciences in Italian universities (36,450 professors).

The index of correlation between the two different rankings of university fields (SDSs) by productivity is over 0.98 in 50% of cases, and is never less than 0.87. Still, in 11 of the 192 SDSs examined we observe cases of universities that undergo shifts of 10 or more positions from one ranking to the next. In the analysis by discipline, the ranking lists are never perfectly overlapping: in every case, in excess of 60% of the universities change position under the different methods. Although the Spearman correlation index is never less than 0.959, various universities experience remarkable changes in position, with jumps of as many as 12 places (seen in Industrial and information engineering and in Mathematics). These variations also affect the "top ranked" universities: with the sole exception of Earth sciences, the best quartile under the two rankings never contains the same universities. Variations in rankings are more noticeable when performance is measured by the *new crown indicator*.

Thus the choice between the two methods of aggregating university performance has a decided influence. The selection must be based on a careful examination of the particular objective for the measure and consideration of the end user's view.

The approach of averaging individual performance detects the distribution of values in the field. This is certainly useful to the research administrator or the student, revealing who it is that conditions the distribution most, in positive and negative senses.

The policy maker, who is instead more interested in maximizing the overall return on research spending, will find it more appropriate to examine the production by the overall institutions and their fields, in relation to the resources employed there. When preparing an assessment exercise for the allocation of resources among universities and their fields, it seems particularly important to differentiate the approach from that used in evaluations for the interests of students and direct administrators. This would maximize the efficiency of the entire allocation process. If the evaluation fails to do this the policy-maker could obtain results that align poorly with their objectives.

There is also the fact that the presence of top scientists has a major impact on universities' performance, as these are the dominant contributors to the output in their fields (Abramo et al., 2013). Where a nation's top scientists tend to be concentrated in a small number of universities, those institutions will invariably appear at the head of the rankings, independent of the method chosen to aggregate performance. However in Italy and other nations featuring non-competitive research systems, the top scientists are instead dispersed among all the universities, along with the unproductive ones (Abramo et al., 2011). This makes the choice of the evaluation methodology particularly stringent.